\documentclass[journal]{IEEEtran}
\usepackage{graphicx}
\usepackage[utf8]{inputenc}
\usepackage{comment}
\usepackage{multirow}
\usepackage{amsmath}
\usepackage[ruled]{algorithm2e}
\LinesNumbered
\usepackage{algpseudocode}
\usepackage{xcolor}
\usepackage{adjustbox}
\usepackage[nameinlink,capitalise]{cleveref}

\begin{document}

\title{Third-Party Hardware IP Assurance against Trojans through Supervised Learning and Post-processing}

\author{
    Pravin Gaikwad,
    Jonathan Cruz,
    Prabuddha Chakraborty,~\IEEEmembership{Graduate Student Member,~IEEE} 
    \\ 
    Swarup Bhunia,~\IEEEmembership{Senior Member,~IEEE} and Tamzidul Hoque,~\IEEEmembership{Member,~IEEE}
\thanks{T. Hoque is with the Department
of Electrical Engineering and Computer Science, University of Kansas, Lawrence,
KS, 66045 USA e-mail: hoque@ku.edu}
\thanks{P. Gaikwad, J. Cruz, P. Chakraborty, and S. Bhunia are with the Department
of Electrical and Computer Engineering, University of Florida, Gainesville,
FL, 32611 USA }}


\maketitle

\begin{abstract}
System-on-chip (SoC) developers increasingly rely on pre-verified hardware intellectual property (IP) blocks acquired from untrusted third-party vendors. These IPs might contain hidden malicious functionalities or hardware Trojans to compromise the security of the fabricated SoCs. 
Recently, supervised machine learning (ML) techniques have shown promising capability in identifying nets of potential Trojans in third party IPs (3PIPs). However, they bring several major challenges. First, they do not guide us to an optimal choice of features that reliably covers diverse classes of Trojans. Second, they require multiple Trojan-free/trusted designs to insert known Trojans and generate a trained model. Even if a set of trusted designs are available for training, the suspect IP could be inherently very different from the set of trusted designs, which may negatively impact the verification outcome. Third, these techniques only identify a set of suspect Trojan nets that require manual intervention to understand the potential threat.   
In this paper, we present VIPR, a systematic machine learning (ML) based trust verification solution for 3PIPs that eliminates the need for trusted designs for training. We present a comprehensive framework, associated algorithms, and a tool flow for obtaining an optimal set of features, training a targeted machine learning model, detecting suspect nets, and identifying Trojan circuitry from the suspect nets. We evaluate the framework on several Trust-Hub Trojan benchmarks and provide a comparative analysis of detection performance across different trained models, selection of features, and post-processing techniques. The proposed post-processing algorithms reduce false positives by up to 92.85\%.   

\end{abstract}
\begin{IEEEkeywords}Trust Verification, Assurance, Hardware Intellectual Property, Golden-free, Machine Learning, Hardware Trojans, Third Party IP, Hardware Trojan Detection, Post-Processing
\end{IEEEkeywords}
\section{Introduction}
System-on-chips (SoCs) today are the backbone of modern electronic devices that are being commonly used in a variety of application domains from defense and aerospace to automotive and the internet of things (IoT). Hardware intellectual property (IP) cores are the basic building blocks of an SoC. The SoC developers often purchase third-party IPs (3PIPs) from different vendors to expedite the development process and cut costs. For instance, along with the dual-core central processing unit, the Apple A9 SoC present in many Apple products incorporates the PowerVR GT7600 graphics processing core from the Imagination Technologies \cite{A9GPU}. 
Several functional verification techniques and tools ensure that the 3PIP core can correctly execute the functions described in the specification. However, SoC integrators cannot verify if the IP is free from unspecified malicious functionalities or hardware Trojans that are designed to stay dormant during functional verification \cite{guo2015pre}. An untrusted IP vendor could introduce hardware Trojans to circumvent the functionality of the SoC or facilitate the leakage of on-chip secret information during field operation. While the malicious alteration of trusted designs at the untrusted foundry has received considerable attention, the corresponding solutions to detect Trojans in ICs do not apply for IP trust verification due to the following reasons:

\begin{itemize}
    \item During procurement, the consumer only receives a high-level specification of the 3PIP from the untrusted IP vendor. There is no golden model or reference implementation to compare the suspect IP.
    \item Application of side-channel analysis-based detection techniques \cite{hoque2017golden, liu2016silicon} is not feasible in this scenario as the reference for parametric behavior (e.g., delay and power) is defined by the untrusted IP vendor itself. \item The lack of reference to the golden model, facilitates the incorporation of large Trojan circuits with complex trigger conditions by an attacker that is hard to activate using logic testing based solutions \cite{saha2015improved, chakraborty2009mero}.
\end{itemize}

Static analysis of the gate-level netlist has shown promising results in identifying malicious nets (wires) by observing one or two specific features (e.g., control metric of the nets
). However, dependency on a very limited number of features allows the attacker to easily redesign the Trojan to bypass them \cite{zhang2014detrust}. The application of machine learning (ML) algorithms allows the incorporation of many features collectively to raise the difficulty of circumventing the countermeasure. Researchers have explored both supervised \cite{hoque2018hardware, hasegawa2016hardware, hasegawa2017trojan} and unsupervised \cite{salmani2016cotd}, \cite{sc_cotd} ML solutions for IP trust verification. One of the essential requirements for applying supervised ML techniques is 
to have a set of known Trojan inserted IPs. 

During the training process of a supervised machine learning model, feature data for each net is extracted from known Trojan inserted IPs, and the nets are labeled (either as Trojan net or normal net) to prepare the data for training
the ML classifiers. These IPs are usually obtained by inserting known Trojans in trusted IPs that do not have any other Trojans. While \cite{hasegawa2016hardware} and \cite{hasegawa2017trojan} relied on known Trojan-inserted IPs available in Trust-HUB, in \cite{hoque2018hardware} the authors suggested that such limited number of custom IPs may not be able to represent the diverse design space for any given class of Trojan (e.g., combinational, sequential, always-on). Instead, an automatic Trojan insertion tool was used to insert a large number of diverse implementations for each class of Trojans in various Trojan-free or trusted IPs \cite{hoque2018hardware}. 

Even if a set of IPs (e.g., signal processing, crypto, communication cores) are available that are guaranteed to be trusted, the suspect IP (e.g., a microprocessor) could be inherently very different from trusted IPs with respect to the behaviour of their internal nets. This discrepancy in implementation among the training and suspect IP could reflect in their corresponding feature data of nets, and could negatively impact the verification outcome.

ML-based techniques have proposed various structural and functional features of a net. But, there are no solutions for identifying the best subset of features for different broad classes of Trojans. For ML based techniques, feature selection could accelerate the verification process by reducing the required time for feature data extraction and improving detection accuracy. Additionally, all of these features are defined assuming the design with no-scan assumption. In practice, while testing the design, the tester can access all the inputs and outputs of a flip-flop with the full-scan assumption of the design. Hence, if an attacker constructs the Trojan with a full-scan assumption, the chances of detecting the Trojan with a no-scan assumption are very low as the inserted Trojan will get very rarely triggered. Another drawback of most of the existing static analysis-based Trojan detection solutions (including the ML-guided techniques) is that instead of detecting malicious circuits, they only report a set of potential Trojan nets in a suspect IP. Thus manual intervention is required to do further analysis on the reported suspicious set of nets. Representing the verification results as a set of suspicious nets impedes understanding of the potential threat and weakens the confidence in the verification result. Such representation does not report the number of Trojans these nets may constitute, their trigger condition, and activation impact. Without this information, it is hard to make a definitive claim regarding the trustworthiness of an IP.

In this paper, we propose \textbf{VIPR} (\underline{V}erification of \underline{IP} T\underline{R}ust), a novel approach to leverage supervised ML for trust verification that does not require a set of trusted sample designs for generating the training data.
To eliminate the need of trusted designs, we propose a mechanism to utilize the untrusted IP itself for evaluating the trust.
During the training phase, we label all the nets in the suspect IP as benign nets (i.e., not part of any Trojan). Next, we insert a large number of varied Trojans in the suspect IP using an automated Trojan insertion tool and label the nets of the resulting synthetic Trojans as Trojan nets. We extract the feature data for these benign and Trojan nets and then train a machine learning classifier. We assume that the attacker is very likely to insert very few Trojans (compared to what we insert using the Trojan insertion tool). Therefore, the training set contains those few data points of attacker-inserted Trojan nets mislabeled as benign nets. However, by incorporating a significantly larger population of tool-generated Trojan data points labeled as Trojan nets, we eliminate the impact of those few mislabeled data points during the training. In other words, by introducing a large amount of correctly labeled Trojan instances in the training data, we ``overpower'' the influence of a few mislabeled data points over the training process \cite{self_train_w_noisy_data}. Our proposed framework also systematically identifies the best combination of features for detecting different classes of Trojans. 
 Furthermore, along with reporting a set of suspicious nets, our approach goes one step further and identifies malicious circuits with distinct trigger and payload logic.
Once the ML model identifies a set of suspicious nets in the 3PIP, the nets are analyzed using an algorithm to construct the Trojan circuit(s). In particular, in this paper, we make the following major contributions:

\begin{itemize}
    \item We propose a golden-free supervised ML-based third-party IP trust verification framework.  Contrary to state-of-the-art techniques, our assumptions are practical and eliminate the need for trusted designs for training. 
    
    \item We incorporate an automated design-specific feature selection flow for generating lightweight ML models and improved classification performance. We present a complete tool flow for the proposed verification process.
    
    \item We propose a set of parameterized post processing algorithms for localizing and reconstructing the hardware Trojan based on the output of the ML model.  The proposed method reduces the need for manual inspection of the suspect nets and reduces false detection.  
    
    \item We extensively verify the efficacy of the proposed framework through a series of well defined quantitative and qualitative analyses on various Trust-Hub Trojans.
    
\end{itemize}

The rest of the paper is organized as the following: Section II describes the existing solutions and their drawbacks; Section III describes the proposed methodology; Section IV  presents results on the Trojan detection performance of our proposed approach, and finally, we conclude in section V.

\section{Background and Related Works}

\begin{figure*}[h]
	\centering
    \includegraphics[width=\textwidth]{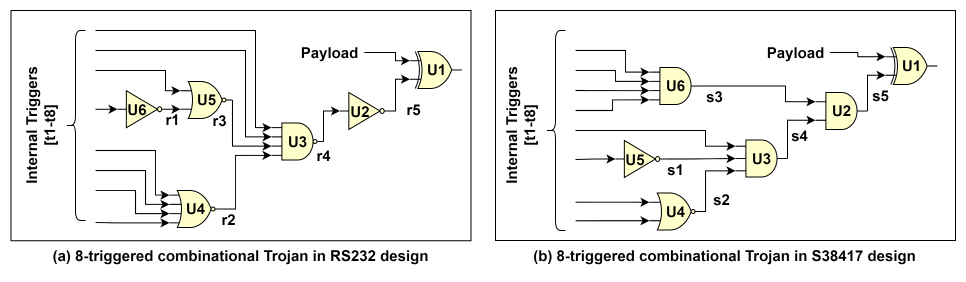}
	\caption{(a) Random tool inserted 8-triggered combinational Trojan in suspect RS232 design. (b) Random tool inserted 8-triggered combinational Trojan in suspect S38417 design.} 
	\label{fig:psr_need}
	
\end{figure*}

\subsection{State-of-the-art Hardware Trojan detection techniques}
Trojan detection in IC/IP is a long standing problem. Several techniques have been proposed over the years to detect Trojans in ICs through machine learning based power side channel analysis \cite{side_channel_pc, untrusted_cots}. Works like ~\cite{signed} use watermarking techniques to detect Trojan behaviour. Authors in ~\cite{signed}, sample critical nets in the design to generate a unique signature. A modification to the original design through the addition of extra gates would cause the switching activity to be modified and thus generate a different signature.

On the other hand, existing methods for hardware Trojan detection for IPs (i.e., gate-level/RTL codes) can be categorized into dynamic and static analysis-based techniques. In the case of dynamic methods, a targeted set of test vectors are generated to activate and detect potential hardware Trojans via simulation \cite{MERO} \cite{MERS} \cite{scalableMERS}. These techniques require exploring the trigger states to activate the Trojans and observe their malicious impact. However, with little to no knowledge of the possible locations of the triggers in a design, the test pattern generation problem becomes difficult \cite{testPatDifficulty}. For static approaches, structural and functional properties related to Trojans are used for detection \cite{hasegawa2016hardware}. Static techniques can be further categorized into search-based, threshold-based, and machine learning-based methods \cite{survey_1}. For search-based methods presented in \cite{UCI} and \cite{VeriTrust}, it is hard to find Trojan nets in very large designs. In \cite{UCI}, the approach relies on test pattern simulation to determine a set of unused nets in the design. Here, the result will depend on a set of patterns used during the simulation, and the test pattern set might not be able to cover all possible combinations of test patterns. The method proposed in \cite{VeriTrust} is only applicable to combinational designs. Functional analysis for nearly unused circuit identification (FANCI) presented in \cite{waksman2013fanci} uses Boolean function analysis, which is a threshold-based static approach to detect hardware Trojans in untrusted IPs. This work is extended by adding graph neighborhood analysis in \cite{HAL}. Because of the time complexity involved in performing the Boolean analysis for large fan-in logic, it is difficult to use \cite{waksman2013fanci, HAL} in large designs \cite{hasegawa2017trojan}.

\begin{table}[t]
\centering
\caption{Feature values of trigger enable signals in RS232 and S38417 design}
\label{tab:psr_features}
\scriptsize\addtolength{\tabcolsep}{8pt}
\begin{tabular}{|c|c|c|c|}
\hline
\textbf{\#}  &   \textbf{Features} &  \textbf{r5 in Fig.(a)} &    \textbf{s5 in Fig.~\ref{fig:psr_need}(b)} \\  \hline
1           &   Probability         & $2*10^{-6}$       & $1*10^{-9}$   \\ \hline
2           &   0-controllability   & 1                 & 1             \\ \hline
3           &   1-controllability   & 43                & 271           \\ \hline
4           &   Observability       & 2                 & 2             \\ \hline
\end{tabular}
\end{table}

Recently, researchers have started exploring supervised and unsupervised machine learning techniques for identifying suspect Trojan nets in untrusted 3PIPs. Table \ref{tab:existing} presents recent efforts using ML for Trojan detection and their limitations. Unsupervised machine learning presented in \cite{salmani2016cotd}, uses combinational Sandia Controllability and Observability (SCOAP) features to cluster the nets in a design and potentially identify Trojan signals which are difficult to control and observe. Here, they have used the TetraMax tool, which limits the SCOAP to a maximum value of 254. 
In practical designs, SCOAP values can go beyond this preset value; hence this technique may produce multiple false positives in the classification of designs. 

In \cite{hasegawa2016hardware}, Hasegawa et al. use supervised support-vector based machine learning model to classify Trojans in suspicious designs. Here, different publicly available sets of Trojan designs are analyzed. However, the authors use the same limited number of Trust-Hub benchmarks for training, which does not contain a large number of diverse implementations of Trojans. Moreover, the authors tested the model using leave-one-out cross-validation, where the suspect IP being tested is also in the training data with all the non-Trojan nets labeled correctly. 
This work is extended by adding a few more gate-level features, and multi-layer neural networks is used for classification in \cite{nn_ht} and random forest classifiers is used in \cite{hasegawa2017trojan}. 

In \cite{sup_scoap}, authors have used SCOAP features to train supervised machine learning models. Here, the training dataset is created synthetically by oversampling the Trojan class. They have used an open-source `Testability Measurement Tool,' which evaluates SCOAP features for designs provided in bench (.bench) format. An infinite value is returned for the nets in a loop, which will impact the results as most sequential designs contain some form of loop.

\begin{table*}[t]
\centering
\caption{Qualitative Comparison of the Proposed Approach with other ML-based Trust Verification Techniques}
\label{tab:existing}
\scriptsize\addtolength{\tabcolsep}{5pt}
\begin{tabular}{|c|c|c|c|c|c|c|c|}
\hline
\multicolumn{1}{|c|}{\textbf{Properties}} 
    & \begin{tabular}[c]{@{}c@{}}\textbf{Hoque}\\ \textbf{et al. \cite{hoque2018hardware}} \end{tabular} 
    & \begin{tabular}[c]{@{}c@{}}\textbf{Salmani }\\ \textbf{et al. \cite{salmani2016cotd}}  \end{tabular} 
    & \begin{tabular}[c]{@{}c@{}} \textbf{Hasegawa } \\ \textbf{et al. \cite{hasegawa2016hardware,nn_ht}} \end{tabular} 
    & \begin{tabular}[c]{@{}c@{}}\textbf{SC-COTD }\\ \textbf{et al. \cite{sc_cotd}} \end{tabular} 
    & \begin{tabular}[c]{@{}c@{}}\textbf{Yang}\\ \textbf{et al. \cite{side_channel_pc}}  \end{tabular} 
    & \begin{tabular}[c]{@{}c@{}}\textbf{Yang }\\ \textbf{et al. \cite{untrusted_cots}}  \end{tabular} 
    & \textbf{Proposed} \\ \hline

Detection Type                                  & Supervised    & Unsupervised  & Supervised    & Unsupervised  & Supervised & Unsupervised & Supervised    \\ \hline                 
Golden Design Required        & Yes            & Yes           & Yes            & Yes           & Yes & No &   No       \\ \hline
Trojan Localization Done                            & No            & No            & No            & No            & No & No &   Yes      \\ \hline
Class-wise Training                             & Yes           & No            & No            & No            & Yes & No &   Yes       \\ \hline
Post-processing                                 & No            & No            & No            & No            & No & No &   Yes     \\ \hline
{Targeted Feature Selection}                    & No            & No            & No            & No            & No & No &   Yes     \\ \hline
\end{tabular}
\end{table*}

\subsection{Major Challenges in ML-guided Solutions}

\subsubsection{Need to learn the complete Trojan space}
To have some form of understanding about the hardware Trojan space, some hardware Trojan benchmarks are made publicly available on Trust-Hub. But the total number of Trojan circuits are still limited in number. Hence, there is a need to have more data on the number of valid Trojan circuits that are hard-to-activate, valid, and have different Trojan structures. 

Also, for any supervised ML technique to work, there is a need to learn as much as possible about the feature space for ML models to derive the hyper-plane that can separate different types of class data.

\subsubsection{Requirement to learn the design specific bias}
Many researchers have presented various works to detect hardware Trojans with the help of various machine learning techniques. Authors in \cite{hoque2018hardware} proposed a method to generate the training database by relying on Trojan free IPs that are publicly available. They have generated multiple Trojan inserted designs with the help of the tool presented in \cite{cruz2018automated} and used various ML models to train the classifier. The disadvantage of this method is that the ML model may learn information specific to the training benchmarks, which may hamper the ML accuracy on the test designs. 
Fig.~\ref{fig:psr_need} shows random 8-trigger combinational Trojan template inserted in designs RS232 (Fig.~\ref{fig:psr_need}(a)) and S38417 (Fig.~\ref{fig:psr_need}(b)). Here, the designs are different structurally as well as functionally. Table \ref{tab:psr_features}, shows the static proabability of net being at logic 1, combinational controllability of 0, combinational controllability of 1 and combinational observability feature values of the final trigger enable signals (r5 and s5) for these two designs. For both the trigger enable signals, it is hard to transition the signals to logic level 1, which can be observed from their probability and combinational controllability of 1 values. Additionally, we can observe that the Trojan in S38417 is harder to activate than the similar Trojan in the RS232 design. Though the previous approach is correct in creating a valid training database, it is not necessarily optimized for the suspicious design under test. The Trojan space is already large enough -- learning both design and Trojan space would require a significant amount of data. If the ML model can learn how the Trojan feature space of a suspicious design behaves then we can reduce the amount of required data for training while also tailoring the ML classification performance to the suspicious design under test. Hence, there is a need to take advantage of the feature space of suspicious design to make design bias decisions.

\subsubsection{Towards lightweight ML model with feature selection}
In literature, various features are proposed related to hardware Trojans based on what an attacker can use to insert the malicious circuit in an IP. However, feature relevance is dependent upon several factors such as type of Trojan, type of benchmark. 
With the increase in the number of features, the feature space explodes, making training more challenging. In practice, enough Trojan inserted designs are not available in the public domain. Additionally, machine learning models are agnostic to domain knowledge. Hence, there is need to reduce the feature space.

\subsubsection{Automated post-processing}

For existing supervised and unsupervised hardware Trojan detection methods presented in the literature, predictions from the ML model are considered as the final result. In a practical situation, predictions of a net to a Trojan class are not meaningful until manual intervention is performed to verify the final predictions. Moreover, the cases in which the ML model outputs a large number of false-positive predictions, manual intervention will be time-consuming and potentially lead to erroneous conclusions for identifying the Trojan circuit. Hence, there is need to automate the post-processing such that the end result will be a reconstructed Trojan structure after removal of some of the false positives.

\section{Methodology}
\begin{figure*}[!h]
	\centering
    \includegraphics[width=.9\textwidth]{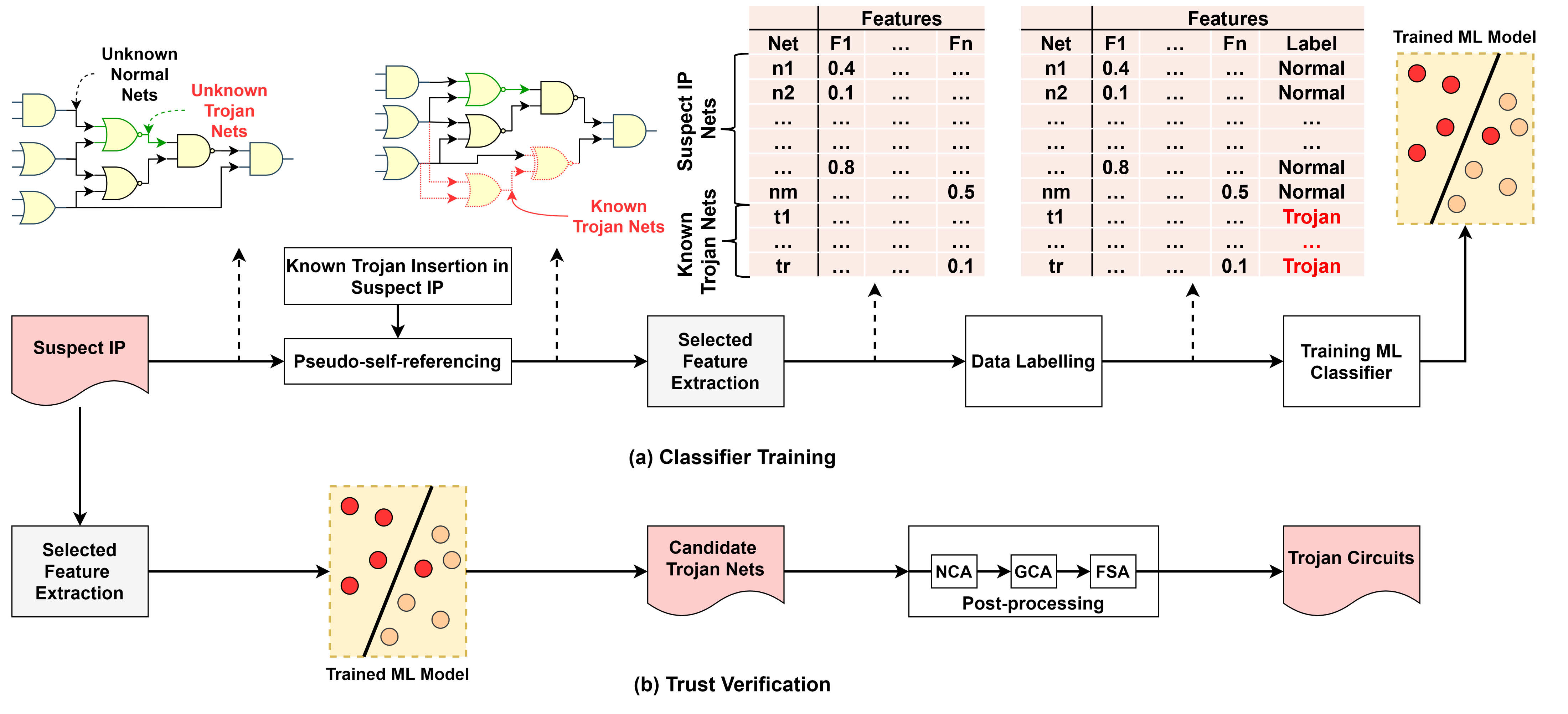}
	\caption{The flow of our proposed framework: (a) a large number of Trojans are inserted in the suspect IP, and the feature data is extracted from them. The labeling process marks all nets of the suspect IP as `normal net' and the nets of our inserted Trojans as `Trojan net'. A trained ML model is generated from this data. (b) During verification, the feature data of suspect IP's net are provided as the test data to the trained model. The trained model identifies the Trojan nets from which the malicious circuits are constructed.} 
	\label{fig:PSR}
	
\end{figure*}
\subsection{Framework Overview}
The overall flow of our proposed approach is presented in Fig.~\ref{fig:PSR}. 
The primary component of our method is a trained ML model that is capable of categorizing the nets of the suspect IP by observing their corresponding feature data. Therefore, as shown in Fig.~\ref{fig:PSR}(a), the first phase of the overall process involves training the ML classifier using a novel technique we propose. In contrast to the training process of earlier supervised ML techniques \cite{hoque2018hardware}, the proposed training method uses the suspect IP itself for generating the training data instead of relying on a set of trusted designs. Previous works on self-referencing have shown techniques to use untrusted hardware components as the reference to detect hardware Trojans in integrated circuits (ICs) \cite{hoque2017golden, du2010self}. We identify our training method as pseudo-self-referencing (PSR) based as introduced conceptually in \cite{chakraborty2021sail}. Our framework uses the untrusted IP itself for training after inserting a large number of known Trojans in it. To evaluate the trained model, a selected set of features are extracted from the suspicious design, and a trained model is used to classify the nets. Finally, post-processing algorithms are applied to extract the possible Trojan circuitry, as shown in Fig.~\ref{fig:PSR}(b).

The training and verification process is executed separately for each broad class of Trojans (e.g., combinational and sequential). 
The PSR-based training process starts with the suspect IP, which is in its gate-level representation. For the desired class of Trojan to be detected, a large number of diverse Trojan implementations of that class are inserted into the suspect IP. To automate this process, we follow the tool-based insertion technique presented in \cite{cruz2018automated}. Next, we extract a large number of functional and structural features for each net of the Trojan inserted suspect IP. The nets are then inspected for labeling. Since we do not know if any of the nets of the suspect IP are malicious, we label all of them as normal nets. The nets of the known Trojans inserted by our automated Trojan benchmarking tool are labeled as Trojan nets. This labeling process facilitates PSR-based training. 
In training data, nets of the original suspicious designs are considered only once as these nets get repeated multiple times with little to no variation in feature values. Our assumption is that if the number of correctly labeled Trojan nets introduced though PSR is significantly higher compared to the number of Trojan nets in the suspect IP that has been labeled as normal net, the classifier will consider the few mislabeled nets as an outlier \cite{self_train_w_noisy_data}. 
We train our preferred ML algorithm using this labeled data and generate the trained model capable of detecting a specific class of Trojan.

The second phase of the overall process is verifying the suspect IP using the trained model. As shown in Fig.~\ref{fig:PSR}(b), all the features are extracted from the suspect IP that was originally obtained from the untrusted vendor. This suspect design is applied as the test data to the trained ML model obtained earlier. The ML model categorizes each net of the suspect IP either as a Trojan net or a normal net. The identified Trojan nets are further processed using a post-processing routine to construct the malicious circuit.

\subsection{Pseudo Self Referencing (being Golden Free)}
To utilize an untrusted IP for training, a large number of Trojans of varied implementation must be generated and inserted in it. While Trust-Hub currently contains examples of different classes of Trojans applicable to gate-level designs, they cover a very small segment of the possible design space for any given class of Trojan. If the construction of the Trojan to be detected is analogous to the design space that is uncovered in the training data, the trained model may not be able to detect it. Hence, it is critical to cover the design space of stealthy Trojans that is undetectable by functional verification. 

\begin{figure}[h]
    \centering
    \includegraphics[width=0.9\columnwidth]{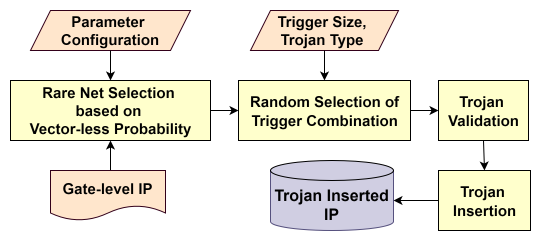}
    \caption{Automatic Trojan Insertion Framework used to insert broad classes of Trojans to generate the training data.}
    \label{fig:trit_flow}
\end{figure}

To automate the Trojan insertion process, we implement the tool-based Trojan insertion technique presented in \cite{cruz2018automated} as shown in Fig.~\ref{fig:trit_flow}. 
The tool automatically generates hardware Trojans and inserts them in appropriate location of the IP such that they cannot be easily detected during functional verification. The tool also verifies using formal tools that the Trojans are not ``dead logic" and can achieve their rare activation values from the primary inputs. Several features of the framework determine the location and the structure of trigger and payload of the Trojan, such as the number of triggers, rarity of the triggers, and inclusion of non-rare triggers. The tool can also insert any number of Trojans in the design.

For inserting hard-to-activate combinational and sequential classes of Trojans we select and combine nets with low activation probabilities. To estimate the signal probabilities of the gate-level IP, vector-less based approach is used as simulating the design with a large number of input vectors might not give accurate or stable results.
Both probabilities of a net being at a logic value 0 ($P_0$) and the probability of a net being at logic value 1 ($P_1$) are calculated for each net by looking at the driving gate of the respective nets. By analyzing $P_{0,1}$ for all nets of the IP, candidate locations for Trojan insertion are identified such that the $P_0$ or $P_1$ is less than a predetermined probability threshold. The tool then selects observable nets as payloads while ensuring no combinational loops are formed. Once the locations are identified, we validate the Trojans using formal tools. Once, valid set of trigger nets and payload nets are obtained, combinational or sequential Trojans are constructed and inserted in the suspicious design. In the case of combinational Trojan, the Trojan structure as well as gates used in the Trojan body varies across different sets of Trojan. For sequential Trojans, we generate finite state machines with different transitions and numbers of states or simple counters with different maximum values.

\subsection{Feature Extraction Methodology}
We extract two different classes of features for training the ML model: functional and structural features. To derive the features from each design, a hyper-graph is created from the gate-level netlist, and different types of algorithms are applied based on the feature. In the following section, we describe all the features used in our framework.

\begin{table}[t!]
\centering
\caption{List of Features used in the Proposed Method.}
\label{tab:func}
\scriptsize\addtolength{\tabcolsep}{0pt}
\begin{tabular}{|l|l|p{5 cm}|} \hline
 \textbf{\#} & 
 \multicolumn{1}{|c|}{\textbf{Functional Feature}} & 
 \multicolumn{1}{|c|}{\textbf{Description}}                         \\ \hline
1   & Static Probability        & Static probability of the net.    \\  \hline
2   & Transition Probability    & Activity from 0 to 1.             \\  \hline
3   & Controllability           & Controllability of the net.       \\  \hline
4   & Observability             & Observability of the net.         \\  \hline
5 & Fanin Level 1 & \# of connected inputs at level 1    \\ \hline
6 & Fanout Level 1 & \# of connected outputs at level 1   \\ \hline
7 & Fanin Level 2  & \# of connected inputs at level 2 \\ \hline
8 & Fanout Level 2  & \# of connected outputs at level 2  \\ \hline
9 & Nearest\_FF\_D      & Distance of the nearest flip-flop input \\ \hline
10 & Nearest\_FF\_Q    & Distance of the nearest flip-flop output \\ \hline
11 & Min. PI Distance       & Min. distance from nearest primary input \\ \hline
12 & Min. PO Distance      & Min. distance from nearest primary output \\ \hline
\end{tabular}
\end{table}

\subsubsection{Functional Features}
For functional features, the sequential elements are considered as full-scan flip-flops. With this assumption, the outputs of the sequential elements are considered pseudo-primary input, and inputs of the sequential elements are considered pseudo-primary outputs. Finally, hyper-graph is traversed in topological order to evaluate functional features. The functional features that are used in the proposed framework are described below.

\textbf{Static Probability:} The static probability denotes the fraction of time the state of the signal or net is expected to be at logic-1 or logic-0. If static probability (also referred as signal probability) of a net being at logic-1 is 0.4, then it indicates that 40\% of the time the net is expected to be at logic-1. We observed that simulation-based probability calculation gets significantly affected by the types of test vectors and the number of total vectors. Hence, we have used a vector-less approach to derive the signal probability of each net in the design. 

\begin{table}[h]
\centering
\caption{Static Probability of basic gates \cite{stat_prob}}
\label{tab:stat_prob}
\scriptsize\addtolength{\tabcolsep}{18pt}
\begin{tabular}{|c|l|l|}  \hline
 \textbf{\#} & \multicolumn{1}{|c|}{\textbf{Gate}} & \multicolumn{1}{|c|}{\textbf{Signal Probability of 1}}                \\ \hline
1          & NOT    & {$1 - P_A$}        \\  \hline
2          & AND    & {$P_A*P_B$}        \\  \hline
3          & OR     & {$P_A + P_B - P_A*P_B$}        \\  \hline
\end{tabular}
\end{table}

The vector-less probability feature calculations are an excellent compromise between accuracy and compute efficiency compared to the simulation-based calculations that are done for a reasonably long duration with realistic stimulus. Equations of basic gates from \cref{tab:stat_prob} are referred to extract static probability feature of all the nets while traversing the hyper-graph.  

\begin{table}[h]
\centering
\caption{Transition Probability of basic gates \cite{trans_prob}}
\label{tab:trans_prob2}
\scriptsize\addtolength{\tabcolsep}{3pt}
\begin{tabular}{|c|l|l|}  \hline
 \textbf{\#} & 
 {\textbf{Gate}} & 
 {\textbf{ \begin{math} P_{0 to 1} = P_{out=0} * P_{out=1} \end{math}}}     \\ \hline
1   & NOT    & {\(1-A\)*\(A\)}                                           \\  \hline
2   & AND    & {(\(1-A*B\))*(\(A*B\))}                                   \\  \hline
3   & OR     & {\begin{math}(1-A)*(1-B)*(1-(1-A)*(1-B))\end{math}}                   \\  \hline
4   & NAND   & {\begin{math}(A*B)*(1-A*B)\end{math}}                                 \\  \hline
5   & NOR    & {\begin{math}(1-(1-A)*(1-B))*(1-A)*(1-B)\end{math}}                   \\  \hline
6   & XOR    & {\begin{math}(1-(A + B - 2*A*B))*(A + B - 2*A*B)\end{math}} \\  \hline
\end{tabular}
\end{table}

 \textbf{Transition Probability:} 
The transition probability (referred to as toggle rate or activity) is the number of transitions from logical level 0 to 1 or 1 to 0. Transition probability can be obtained by either simulating set of test vectors or with a vector-less approach. In the simulation approach, the transition probability gets affected by the order in which vectors are passed while performing the simulation. For instance, if two consecutive vectors used during simulation differ by only one bit, then the chances of observing transition on the nets affected by this bit change are low. Hence, for the transition probability feature as well, we have used the vector-less approach to derive transition probability for each net in the design, which considers the type of gate driving a particular net to derive the feature value. \cref{tab:trans_prob2} shows the equation of basic gates that are used to extract the transition probability of each of the net while traversing the hyper-graph from inputs to outputs.

\textbf{Controllability and Observability:} The controllability metric captures the effort in assigning a net to the desired logic value by applying vectors to the primary inputs. Observability represents the effort of propagating the logic state of a net to observable points (e.g., primary outputs). Goldenstein developed the SCOAP (Sandia Controllability and Observability Program) testability measures \cite{scoap_sandia}. The SCOAP value can range from zero to infinity. The higher the value, the more effort is required to control or observe a signal in the design. For each signal in the design, there will be six numerical values, i.e., combinational and sequential variants for 0-controllability, 1-controllability, and observability. Combinational controllability measures the effort required in achieving  logic 0/1 of a signal via the primary inputs whereas sequential controllability measures this effort via the amount of sequential elements that must be clocked. 
For observability, combinational observability is the measure of the effort required to observe the signal value at an observable point (i.e.: primary output, scan-flop), while sequential observability is the measure of the amount of sequential elements that must be clocked to observe the signal. SCOAP values are heavily influenced by the presence of test infrastructure available in the design. To this end, we have implemented two variations of the SCOAP measure. The first variant assumes the every sequential element is part of a scan-chain (full-scan), we have obtained combinational SCOAP values for each of the nets in the gate-level design in the first approach. Here, the combinational controllability is obtained by traversing the hyper-graph from inputs to outputs while evaluating the features based on controllability equations of basic gates. Similarly, for combinational observability, the hyper-graph is traversed in reversed order and basic gate equations are used to evaluate features of individual nets. In the second variant, we have extracted the SCOAP measure assuming no sequential element belongs to a scan-chain (no-scan) as per \cite{scoap_sandia}, \cite{sc_cotd}. 

Some of the previous works on hardware Trojan detection rely on Synopsys's TetraMax tool to extract the SCOAP measure. One of the limitations of the TetraMax tool is that the maximum value that the tool can assign to a net is limited to 254. So for nets that can have values greater than 254 get assigned with an asterisk symbol. Hence, we have not used the TetraMax tool to extract this feature as the SCOAP values can go beyond the set threshold present in TetraMax. Publicly available ``Testability Measurement Tool" does not support designs in Verilog format, and for designs that consist of loops, an infinity value is returned. 
Hence, we have implemented the algorithm from scratch to evaluate the SCOAP features of the design. 
To design a Trojan that is hard to activate during logic testing, an attacker is expected to design the trigger mechanism using nets with low controllability (i.e., higher controllability measure). To further hide the Trojan, the corresponding payload could be designed to impact only low observable (i.e., high observability measure) nodes. These two features have been used for detecting Trojans using unsupervised clustering \cite{salmani2016cotd}.

\subsubsection{Structural Features}
\begin{figure}[!th]
	\centering
	\includegraphics[width=\columnwidth]{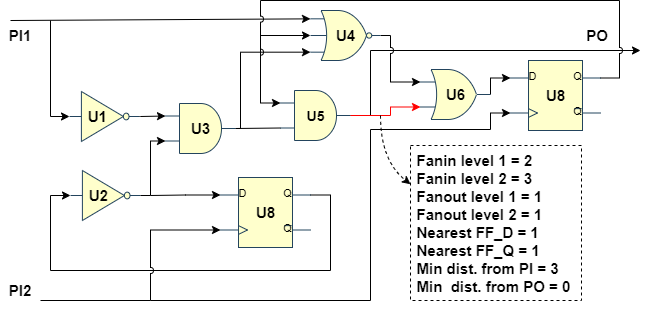}
	\caption{Illustration of structural features for the net marked in red. } 
	\label{fig:struct}
	
\end{figure}

Structural features describe the location and connectivity information of the net. While most of the structural features may appear uncorrelated in categorizing a Trojan net, they provide useful information when used together with functional features \cite{hoque2018hardware}.
Table \ref{tab:func} lists all the structural features (5 to 12) that were analyzed in the paper, with associated illustration in Fig.~\ref{fig:struct}, and described below.

\textbf{Fanin and Fanout:} Fanin is useful in understanding if a net is driven by a logic with a large fanin structure. 
Level 1 fanin of a net represent the number of nets that are input to the cell driving the net. Level 2 fanin indicates the total fanin of those nets that are Level 1 input to the driving cell. Level 1 fanout represents the number of cells the net is propagating to. The trojan circuit usually provides output to the original design through the payload only. Fanout of gates in the logic cone of a trigger circuit usually goes as inputs to other gates in the trigger logic. Hence, the fanout of Trojan nets is likely to be low. Even for always-on Trojans using ring-oscillators or shift-registers, the fanout of the nets does not propagate to the normal part of the design \cite{hoque2018hardware}.

\textbf{Distance from nearest Flip-Flop:} We consider combinational and sequential Trojans as two different classes of Trojans and separate their training and testing process. Since sequential Trojans nets are expected to have flip-flops in close locations, unlike the combinational Trojan nets, we extracted the distance from the nearest flop-flop as a possible feature (i.e., Nearest\_FF\_D and Nearest\_FF\_Q). While these features should help a trained model for sequential Trojans to detect sequential Trojan nets in a suspect IP, it also should prevent the model from classifying combinational Trojan nets in the suspect IP as nets of sequential Trojans.

\textbf{Distance from Nearest Primary Input and Output:}
Distance from primary input (PI) provides more context to various functional features. For instance, Trojans that take the trigger values directly from the PIs may have higher toggle rates than Trojans that take trigger inputs from very rare signals. However, the toggle rate could be rarer when compared to other non-Trojan nets with similar shorter distances from the PIs. Distance from the primary output (PO) is useful for identifying malicious structures that require them to be situated near the PO (e.g., Trojans that leak information through the PO). Since flip-flops are usually connected to the clock, reset, and enable signals directly coming from the PI, we only consider the D-input when calculating the distance from PI.

\subsection{Design Targeted Feature Selection Methodology}

\begin{figure}[h]
	\centering
    \includegraphics[width=\columnwidth]{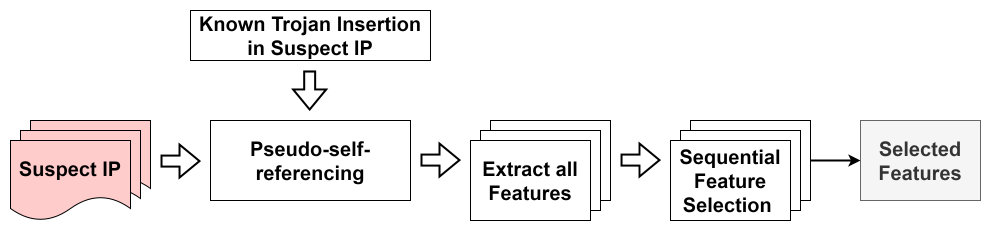}
	\caption{Feature selection flow in the proposed framework.} 
	\label{fig:feature_selection}
	
\end{figure}

Existing studies have explored various structural and functional features of the nets to identify the nets of a Trojan. To train any machine learning model, there is an inherent need of feature space that can separate different class data present in the database.
In \cite{hoque2018hardware}, the authors suggest that combining both structural and functional features has contributed to their improvement in detection accuracy compared to the techniques that use only one type of features \cite{hasegawa2016hardware, hasegawa2017trojan, salmani2016cotd}. However, \cite{hoque2018hardware} implemented the framework in the FPGA netlist, which supports 6 input look-up table (LUT). Features data obtained from FPGA-mapped netlist and gate-level netlist may provide a different data distribution for Trojan and normal nets. A feature useful for classification in the FPGA netlist may not have the same impact in the gate-level netlist. Additionally, to decide which features are relevant, there is a need to filter out some of the features that do not contribute towards learning of the feature space. 
Therefore, our framework includes a feature selection step as shown in Fig.~\ref{fig:feature_selection} to identify the best possible features for detecting Trojans in the suspect gate-level netlist. This technique can also be used in the FPGA netlist. Exhaustive feature selection is performed to find the best possible set of features based on the cross-validation score of a classifier. In each iteration, a feature that gives maximum performance for the given classifier using cross-validation is added to a possible set of features. This step is repeated until we have a set of features that gives the highest accuracy for classification. For each of the suspect designs, a different set of features are reported by the exhaustive search. For further analysis, the top 5 frequently used features across multiple suspect designs are considered to reduce the workload on the machine learning model. Fig.~\ref{fig:train_dist} shows the kernel density estimation plots of these top 5 features for tool inserted Trojan nets and already present Free nets in a suspect design.

\subsection{Localizing Trojans: Post-processing Algorithms}

\begin{figure*}[t]
    \centering
    \includegraphics[width=1\textwidth]{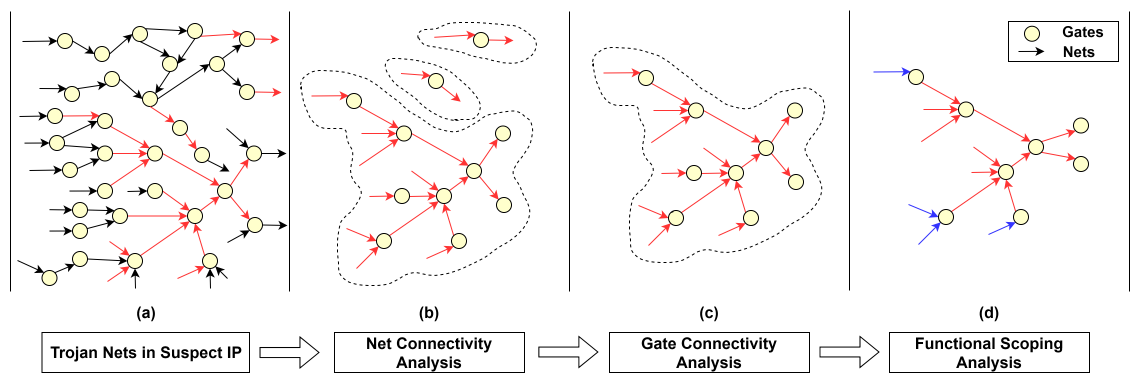}
    \caption{Circuit reconstruction with the proposed post-processing algorithms. Nets highlighted in red color represents predictions from the ML model. Specifically for the last section, nets highlighted in blue are false-positive nets, and those highlighted in red are true-positive nets.}
    \label{fig:post_flow}
\end{figure*}

Each trained model predicts the class of a net present in the suspicious design based on the model decision boundary. Nets are either classified as \emph{Trojan} or \emph{Free} nets. To get more insight on the structure of the Trojan body, Trojan circuit reconstruction is performed with the help of post-processing  \cref{alg:post_algo_1,alg:post_algo_2,alg:post_algo_3}. 

\begin{algorithm}
	\caption{Net Connectivity Analysis (NCA)}
    \label{alg:post_algo_1}
    \SetKwData{Left}{left}\SetKwData{This}{this}\SetKwData{Up}{up}
	\SetKwFunction{Union}{Union}\SetKwFunction{FindCompress}{FindCompress}
	\SetKwInOut{Input}{input}\SetKwInOut{Output}{output}
		\Input{ $d$: Netlist of the suspect design \\ 
		        $t_{nets}$: ML predicted Trojan nets.} 
		\Output{$d_{l1}$: NCA processed suspect design}
	    \textit{$d_{l1}$} = $\emptyset$  \\
		\ForEach{ $currNet \in t_{nets}$}{
	        $fwdNets$ = $fanout(currNet)$   \\
	        \ForEach { $net \in fwdNets$}{
    	        \If{ ($net$ is in  $t_{nets}$) }{
    	           $currNode$ = Driving node of $net$ \\
     	           Add ($currNet$, $currNode$, $net$) to $d_{l1}$    \\
    	        }
    	    }
	        $bwdNets$ = $fanin(currNet)$    \\
	        \ForEach { $net \in bwdNets$}{
    	        \If{ ($net$ is in  $t_{nets}$) }{
    	            $currNode$ = Driving node of $currNet$ \\
    	            Add ($currNet$, $currNode$, $net$) to $d_{l1}$  \\
    	        }
    	    }
	    }
\Return $d_{l1}$ 
\end{algorithm}

Input to the Net Connectivity Analysis (NCA) post-processing algorithm is the suspicious design netlist and predictions from the ML model. The algorithm starts with the construction of a hyper-graph with nodes and edges derived by parsing the netlist. \cref{fig:post_flow}(a) shows a partial hyper-graph of the suspicious design. Here, predictions from the ML model are highlighted with red color nets. As shown in \cref{alg:post_algo_1}, for each of the nets with Trojan class predictions, its immediate fanout and immediate fanin is obtained on line 3 and 10, respectively. These fanout and fanin cones are checked for the presence of another predicted Trojan net, as shown in lines 4-8 and lines 11-16. If Trojan nets are immediately connected to each other through a node, then the node connecting them and the respective edges are added to the $d_{l1}$ graph, as shown in lines 7 and 14. This step removes any of the predicted Trojan nets that are isolated from other Trojan nets. 

Finally, we get the hyper-graph of the ML predicted Trojan nets that are within one level of breadth first search of the other predicted Trojan nets as the output of the NCA post processing algorithm.

\begin{algorithm}
	\caption{Gate Connectivity Analysis (GCA)}
        \label{alg:post_algo_2}
    \SetKwData{Left}{left}\SetKwData{This}{this}\SetKwData{Up}{up}
	\SetKwFunction{Union}{Union}\SetKwFunction{FindCompress}{FindCompress}
	\SetKwInOut{Input}{input}\SetKwInOut{Output}{output}
		\Input{ 
		        $d_{l1}$: NCA processed suspect design \\ 
		        $\theta_{depth}$: Threshold to select connection depth\\
		        $sortedNodes$ : Sorted graph nodes for $d_{l1}$
		       } 
		\Output{$d_{l2}$: GCA processed suspect design}
		\textit{$d_{l2}$} = $d_{l1}$  \\
        \Repeat{ \normalfont $lessConnectedFlag$ $==$ \normalfont 1}
		{
		    \textit{$nodesToRemove$} = $\emptyset$  \\
		    \textit{$lessConnectedFlag$} = 0  \\
		    \ForEach { $node$ in reversed($sortedNodes$)}{
		    
    	        $succHeight$ = $fanout(currNet)$   \\
    	        $predHeight$ = $fanout(currNet)$   \\
    	        \If{ ($succHeight$ $<$ $\theta_{depth}$ and $predHeight$ $<$ $\theta_{depth}$) }{
        	       $lessConnectedFlag$ = 1 \\
        	       Add $node$ to $nodesToRemove$ \\
        	       Remove $node$ from $d_{l2}$ \\
        	    }
    	    }
    	    Remove $nodesToRemove$ from $sortedNodes$ \\
		}
		
     \textbf{return} $d_{l2}$
\end{algorithm}

The NCA processed hyper-graph is passed through Gate Connectivity Analysis (GCA) post-processing algorithm as described in \cref{alg:post_algo_2} to remove nodes that do not have minimum user-provided depth in either forward or backward direction. Additionally, the list of topologically sorted nodes present in $d_{l1}$ is also passed from the NCA post-processing algorithm for further analysis. 

The output of the NCA post-processing algorithm is a disconnected graph with a reduced number of nodes and edges compared to the original hyper-graph of $d$ as shown in \cref{fig:post_flow}(b). Typically, a Trojan body consists of nodes that have some form of connectivity to other nodes present in its neighbourhood. To understand how deeply a node is connected to other nodes, its maximum height is calculated in either direction, as shown on lines 6 and 7. If the node's height is not above the user-provided threshold, the node is removed from the hyper-graph along with its input and output edges. This process is repeated until each of the nodes in the hypergraph has a minimum height in either direction. From \cref{fig:post_flow}(b) and \cref{fig:post_flow}(c), we can observe that the nodes with depth less than user provided threshold gets filtered out after GCA post-processing.  

\begin{equation}
    \label{CC_equation}
    net_{cc} = \sqrt{CC_0^2 + CC_1^2}    \\  
\end{equation}

\begin{equation}
    \label{SC_equation}
    net_{sc} = \sqrt{SC_0^2 + SC_1^2}    \\
\end{equation}


To further eliminate possible false positives, Functional Scoping Analysis (FSA) post-processing is performed as shown in \cref{alg:post_algo_3}. 
Mean value of \cref{CC_equation} and \cref{SC_equation} for Free nets present in the training data is used to get $\lambda_{cc}$ and $\lambda_{sc}$. Here, controllability features with the no-scan assumption are used to see the effect of primary inputs on tool inserted Trojans. For each of the nets present in the hyper-graph, combinational controllability feature ($net_{cc}$), sequential controllability feature ($net_{sc}$) and static probability of net being at 1 ($net_{prob}$) is compared with user provided thresholds as shown in line 5 to add domain specific knowledge regarding hardware Trojans. Finally, nets that fit these criteria are removed from the $TrojanCircuit$ hyper-graph to get the possible Trojan circuit present in the suspicious design as shown in \cref{fig:post_flow}(d).

\begin{algorithm}

	\caption{Functional Scoping Analysis (FSA)}
        \label{alg:post_algo_3}
    \SetKwData{Left}{left}\SetKwData{This}{this}\SetKwData{Up}{up}
	\SetKwFunction{Union}{Union}\SetKwFunction{FindCompress}{FindCompress}
    \SetKwInOut{Input}{input}\SetKwInOut{Output}{output}
		\Input{ 
		        $d_{l2}$: GCA processed suspect design \\ 
		        $\theta_L$: Lower bound on probability \\
		        $\theta_U$: Upper bound on probability \\
		        $\theta_{cc}$: Combinational controllability threshold \\
	        	$\theta_{sc}$: Sequential controllability threshold \\
	        	$\lambda_{cc}$ = Mean combinational controllability of \\
	        	                 Free nets in training data \\
		        $\lambda_{sc}$ = Mean sequential controllability of Free \\
		                nets in training data \\ 
		        } 
		\Output{$TrojanCircuit$: Trojan circuit netlist}
		\textit{$TrojanCircuit$} = $d_{l2}$\\
		\ForEach { $net$ in $Trojan Circuit$}{
	        $\delta_{cc}$ = $net_{cc}$ - $\lambda_{cc}$\\   
	        $\delta_{sc}$ = $net_{sc}$ - $\lambda_{sc}$\\
	        \If{ ($\delta_{cc}$ $<$ $\theta_{cc}$ and $\delta_{sc}$ $<$ $\theta_{sc}$ and $net_{prob}$ $>$ $\theta_L$ and $net_{prob}$ $<$ $\theta_U$ ) }{
    	        Remove $net$ from $TrojanCircuit$ 
    	    }
	    }
     \textbf{return} $TrojanCircuit$
\end{algorithm}

\section{Results}

\begin{figure*}[t]
    \centering
    \includegraphics[width=1\textwidth]{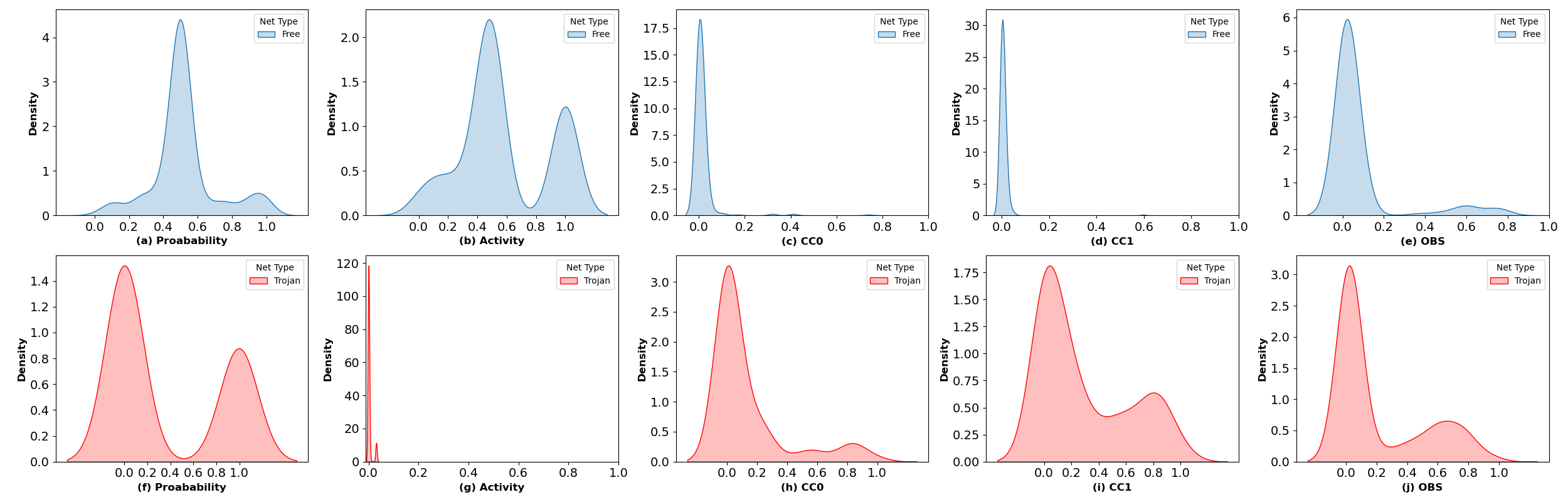}
    \caption{Kernel density estimation plot for MinMax scaled features used in the training of a ML model for RS232-T1400 suspicious design. Here, random valid 50 8-trigger combinational Denial of Service type of Trojans are inserted in the suspicious designs with the help of \cite{cruz2018automated} tool.}
    \label{fig:train_dist}
\end{figure*}

To evaluate the proposed approach, we have used different implementations of sequential and combinational Trojans present in different suspect IPs (RS232 and S38417), which are publicly available on the Trust-Hub site \cite{trust_hub}. Some of these IPs contain either combinational (C) or sequential (S) Trojans denoted by C/S in the first column of Table~\ref{tab:final_voted_results}. The training data is generated from the respective IPs for different triggers sets and Trojan types with the proposed pseudo-self-referencing approach.

\subsection{Feature Selection Analysis}

In our method, we have reduced the set of features required for training a supervised ML model. Forward exhaustive feature selection is performed with the help of a sequential feature selection wrapper available in the open source sklearn library. Probability and activity features are provided as initial features to the sequential feature selection wrapper to ease the exhaustive feature search. These basic features are provided assuming that the attacker should at least consider these minimum set of features while designing a valid hard-to-activate hardware Trojan. A different set of features are selected across different available training databases. We have used the top 5 most frequently used features across all different cases considered while training an ML model. These 5 features turn out to be functional features that relate to the functional behavior of the net based on its driving gate. Table \ref{tab:func} rows 1-4 present the list of functional features that have been incorporated in our framework.

Fig.~\ref{fig:train_dist} shows the distribution of proposed features used in the Training of an ML model for RS232-T1400 design. Here, the suspicious design is used to create the training data by inserting 8-triggered combinational Trojans with the help of an automatic Trojan insertion tool. Fig.~\ref{fig:train_dist} shows kernel density estimation plots of the features scaled using the MinMaxScaler function in sklearn. The plots (a) and (f) show the kernel density estimation of static probability of a net being at logic 1 for Free nets and Trojans nets, respectively. We can see that the inserted Trojans have either very high or very low probability values.  Along with low or high static probability values, we can observe that the activity values are distributed near 0, as shown in plots (b-g). From the distribution of controllability feature of 0 and 1 with full-scan assumption, we can see that some of the Trojans have high controllability values, i.e., they are difficult to control. Similarly, some of the Trojans are hard to observe, as shown in Fig.~\ref{fig:train_dist} (e). All of the observed feature behaviors from \cref{fig:train_dist} provide insight into a Trojan's stealthy behavior.

\begin{figure*}[t]
    \centering
    \includegraphics[width=1\textwidth]{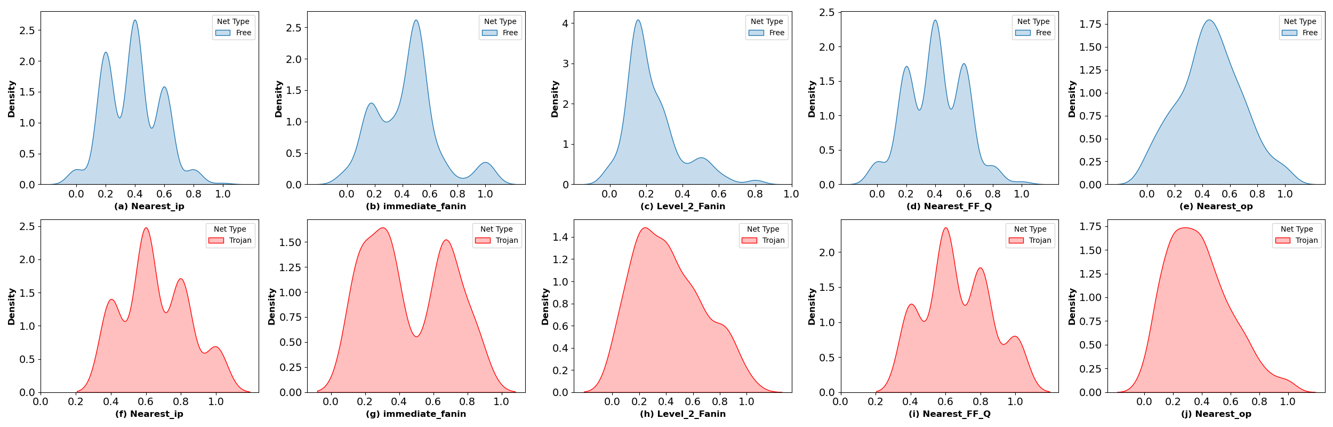}
    \caption{{Kernel density estimation plots of (a) Nearest input, (b) immediate fanin, (c) level-2 fanin and (d) nearest flip-flop Q output (e) nearest output features extracted from 50 8-triggered combinational training dataset of RS232-T1400 suspicious design.}}
    \label{fig:train_surr_dist}
\end{figure*}

Fig.~\ref{fig:train_surr_dist} shows the distribution of some of the structural scaled feature values extracted from the RS232-T1400 training dataset. 
For structural features, we can observe in plot (c) that the Trojan and Free nets have a high density near 0.25 value, and in plot (d), centers for density plots get shifted by a small amount, but there is a large overlap of feature values. Hence, this set of features can't be directly used for the training of an ML model as it will increase the number of false negatives.

\subsection{Detection of Trust-HUB Trojans}

To evaluate the proposed approach, the training database for the ML model is obtained by inserting tool-generated Trojans in the suspicious gate-level design. To select the trigger nets of the Trojan, vector-less probability features of the suspicious design are used by the tool. For each suspicious design, the probability threshold is varied from 0.0001 to 0.1 to select the probable set of trigger nets. ML models are trained based on trigger class as well Trojan type inserted in the suspicious design. Two types of Trojans are inserted in the design, i.e., combinational and sequential Trojans. For each type, 50 instances of 5, 6, and 8 trigger Trojans are inserted in each suspicious UART design. The trigger combination of each Trojan is validated using the Cadence JasperGold tool, and payloads are selected to observe the effect at the primary output. 

\begin{figure*}[t]
	\centering
    \includegraphics[width=.9\linewidth]{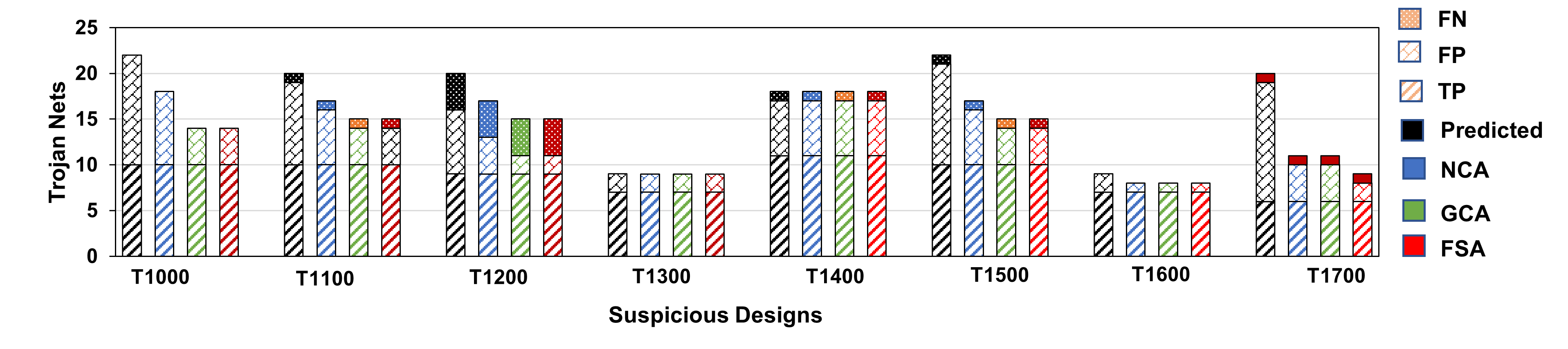}
	\caption{VIPR framework output with structural and functional features for RS232 designs. The ML models are trained with 5, 6, and 8 triggered combinational Trojans. Each of the bars represents voted value of ML predictions followed by three types of post-processing routines for each design.} 
	\label{fig:vipr_framework_merged_comb_func_struct}
\end{figure*}

\begin{figure*}[t]
	\centering
    \includegraphics[width=.9\linewidth]{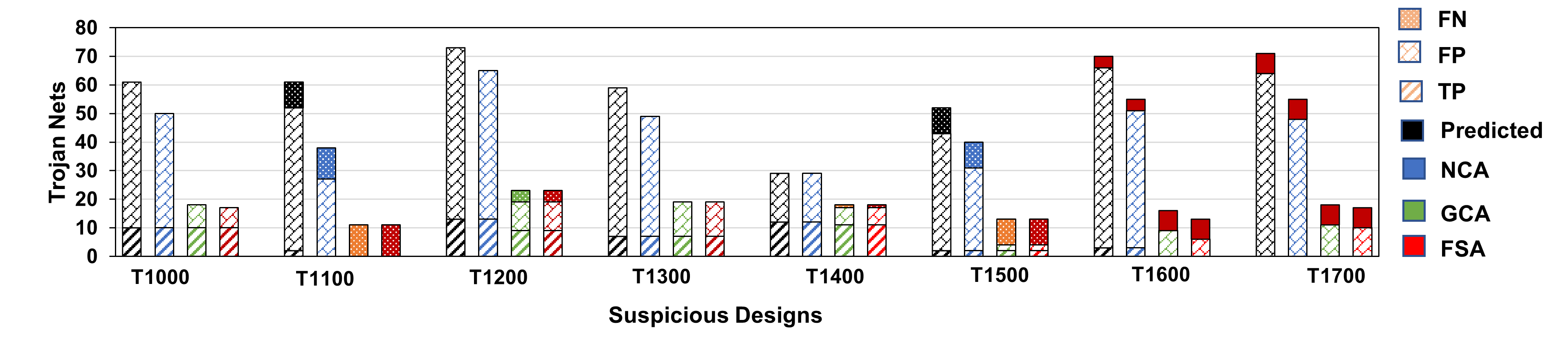}
	\caption{VIPR framework output with structural and functional features for RS232 designs. The ML models are trained with 5, 6, and 8 triggered sequential Trojans. Each of the bars represents voted value of ML predictions followed by three types of post-processing routines for each design.} 
	\label{fig:vipr_framework_merged_seq_func_struct}
\end{figure*}

\begin{figure*}[t]
	\centering
    \includegraphics[width=.9\linewidth]{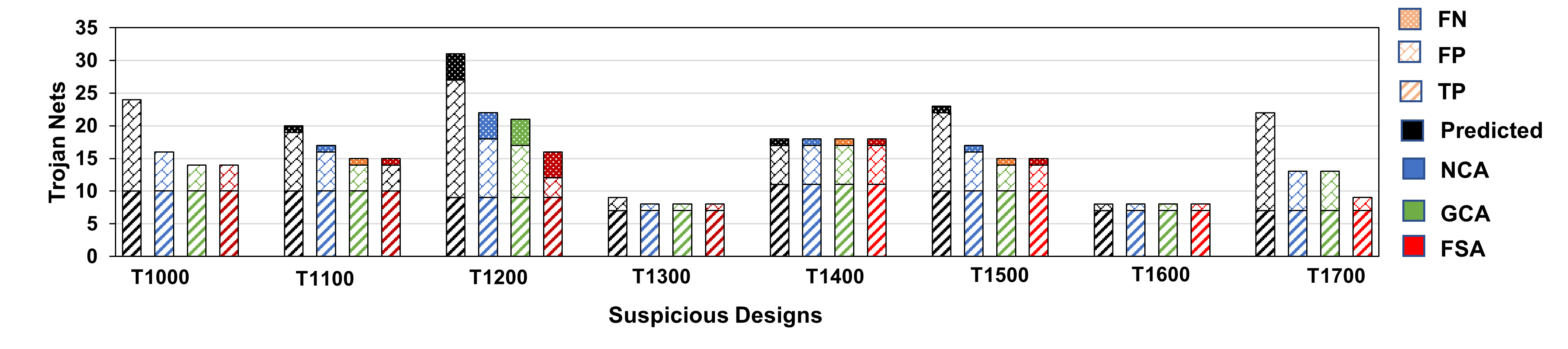}
	\caption{VIPR framework output with selected features for RS232 designs. The ML models are trained with 5, 6, and 8 triggered combinational Trojans. Each of the bars represents voted value of ML predictions followed by three types of post-processing routines for each design.} 
	\label{fig:vipr_framework_merged_comb_func}
\end{figure*}
\begin{figure*}[t!]
	\centering
    \includegraphics[width=.9\linewidth]{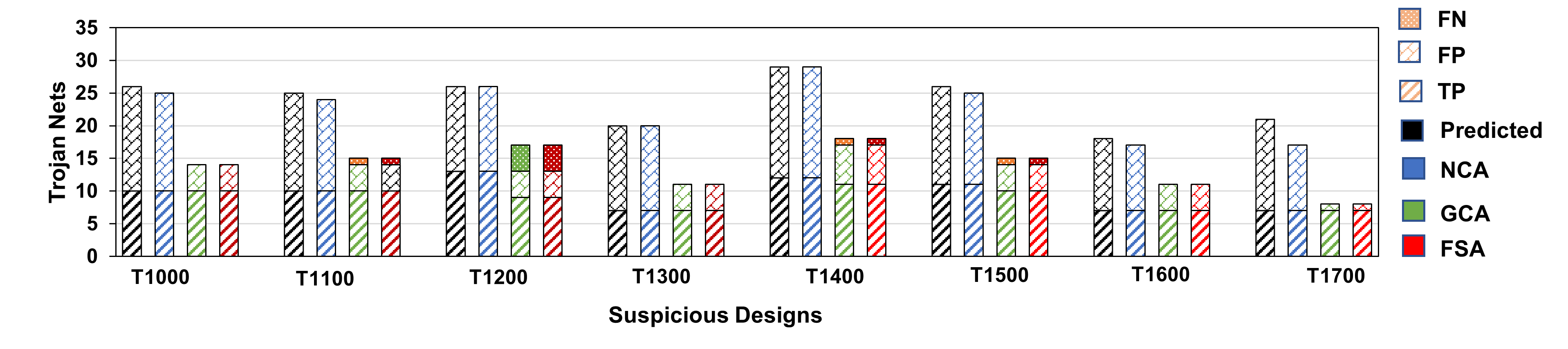}
	\caption{VIPR framework output with selected features for RS232 designs. The ML models are trained with 5, 6, and 8 triggered sequential Trojans. Each of the bars represents voted value of ML predictions followed by three types of post-processing routines for each design.} 
	\label{fig:vipr_framework_merged_seq}
\end{figure*}

All features are extracted from training designs and the suspicious design using the designed feature extraction framework. UART benchmarks from Trust-Hub with 90nm technology nodes are used to test the VIPR framework. Trojan nets in suspicious design are labeled as \emph{Normal} nets in the training data set, and the nets present in the tool generated Trojan are labeled as \emph{Trojan} nets. Both the training and test data are scaled using MinMaxScaler. An SVM model is used to train each ML model based on training class type as it is a binary classification. The hyper-parameters used for the ML model are `rbf' kernel, class\_weight=`balanced', C=0.8. The `balanced' mode is used to automatically adjust the weights of the input data based on the frequency of the class labels. The `C' parameter is chosen to allow very little outliers while performing the classification. To train the model proposed features presented in Table~\ref{tab:func} are used. Testing of each suspicious design is done separately with respective models which are trained on combinational Trojan inserted designs as well as sequential Trojan inserted designs. 
{For GCA post-processing, $\theta_{depth}$ is selected as 2 to find Trojans with at least circuit depth of 2 levels. For FSA post-processing, the lower bound on probability is chosen as 0.4, and the upper bound is selected as 0.6 to eliminate falsely classified Trojan nets whose features belong to the Free feature space.} The following section describes detection accuracy for the test designs after ML predictions and the sequence of post-processing routines.

\begin{table*}[ht]
    \centering
    \caption{VIPR framework final results after majority voting per model class}
    \label{tab:final_voted_results}
    \scriptsize\addtolength{\tabcolsep}{7pt}
    \begin{tabular}{|c|cc|cc|cc|cc|cc|cc|}
    \hline
    \textbf{Suspicious Design}  
                                & \multicolumn{2}{c|}{Comb. training}
                                & \multicolumn{2}{c|}{Seq. training}
                                & \multicolumn{2}{c|}{Comb. + Seq. }  
                                & \multicolumn{2}{c|}{Hoque et al. \cite{hoque2018hardware}}
                                & \multicolumn{2}{c|}{SC-COTD*}
                                & \multicolumn{2}{c|}{SC-COTD \cite{sc_cotd} }\\ 
                                \cline{2-13}
                                & \textbf{FP}   & \textbf{FN}
                                & \textbf{FP}   & \textbf{FN}
                                & \textbf{FP}   & \textbf{FN}
                                & \textbf{FP}   & \textbf{FN}
                                & \textbf{FP}   & \textbf{FN}
                                & \textbf{FP}   & \textbf{FN} \\ \hline
    {RS232-T1000 (C)}       & 4  & 0
                        & 4  & 0
                        & 4  & 0
                        & 4  & 1 
                        & 12 & 4
                        & 2  & 0 \\ \hline
    {RS232-T1300 (C)}       & 1  & 0  
                        & 4  & 0  
                        & 1  & 0  
                        & 6  & 2 
                        & 14 & 2 
                        & 0  & 0 \\ \hline
    {RS232-T1700 (C)}       & 2  & 0 
                        & 1  & 0  
                        & 0  & 0  
                        & 8  & 3 
                        & 0  & 7
                        & NA  & NA \\
                        \hline
    {S38417-T100 (C)}       & 6  & 0 
                        & 6  & 0  
                        & 6  & 0  
                        & NA & NA 
                        & 8  & 1 
                        & 1  & 0 \\
                        \hline
    {S38417-T200 (C)}       & 1  & 0 
                        & 1  & 0  
                        & 1  & 0  
                        & NA & NA 
                        & 0  & 9 
                        & 9  & 0 \\
                        \hline
    {RS232-T1100 (S)}       & 4  & 1  
                        & 4  & 1  
                        & 4  & 1  
                        & 6  & 3 
                        & 12 & 5
                        & 2  & 0 \\
                        \hline
    {RS232-T1200 (S)}       & 3  & 4  
                        & 4  & 4  
                        & 1  & 4  
                        & 7  & 1 
                        & 0  & 11
                        & 2  & 0 \\
                        \hline
    {RS232-T1400 (S)}       & 6  & 1 
                        & 6  & 1  
                        & 6  & 1  
                        & 6  & 0 
                        & 0  & 6
                        & 2  & 0 \\
                        \hline
    {RS232-T1500 (S)}       & 4  & 1 
                        & 4  & 1  
                        & 1  & 1  
                        & 5  & 1 
                        & 12 & 5
                        & 3  & 0 \\
                        \hline 
    {RS232-T1600 (S)}       & 1  & 0 
                        & 4  & 0  
                        & 1  & 0  
                        & NA  & NA  
                        & 2  & 2
                        & 0  & 0 \\
                        \hline                                     
    \end{tabular}
\end{table*}

\cref{fig:vipr_framework_merged_comb_func_struct,fig:vipr_framework_merged_seq_func_struct,fig:vipr_framework_merged_comb_func,fig:vipr_framework_merged_seq} show the final results for individual benchmarks with different set of features resulting from feature selection. The color bars are used to specify the results after a certain stage in the framework, and different patterns are used within each of these bars to show false negatives (FN), false positives (FP), and true positives (TP) for each of the design. For each of the benchmarks, results are represented by 4 bars to show results after ML predictions, NCA, GCA, and FSA post-processing stages in respective order. To summarize the overall performance of the proposed approach at each stage, voting is performed across 5,6, and 8 triggered combinational and sequential types of Trojan training. Voted results are used to calculate the number of false positives, false negatives, and true positives nets. The last bar represents the final result produced by the VIPR framework for each design, which is nothing but the majority voted FSA post-processed nets.

\cref{fig:vipr_framework_merged_comb_func_struct,fig:vipr_framework_merged_seq_func_struct} shows results of the proposed framework without a feature selection process. Here, all the structural and functional features are used while training the ML model. As shown in Fig. \ref{fig:train_surr_dist}, most of the structural features are not useful to distinguish between the feature space of Trojan and normal nets for the respective design. Thus, the ML model is not able to produce a better set of results when all features are considered for training. As all the features extracted for each training class are not useful, there is a need to select the best set of features that can be used to fairly compare the performance of the proposed method across various suspicious designs.

\cref{fig:vipr_framework_merged_comb_func,fig:vipr_framework_merged_seq} show how the VIPR framework with ML model trained on combinational and sequential Trojans reduces false positives with the help of the proposed post-processing routines. Here, a selected set of features are used to derive all the results. We can notice a downward trend for each set of results. In Fig.~\ref{fig:vipr_framework_merged_seq}, we can observe that the number of false positives is relatively high as the types of sequential Trojans used in training are mainly some type of FSMs or counters, whereas publicly available sequential Trojans contains only few flip-flops, which are just used to latch the value in the Trojan body. Particularly for design RS232-T1200 (labeled as T1200), four flip-flops are used to latch the value in the Trojan body. The ML model trained on sequential Trojans can detect these sets of Trigger nets driven by flip-flops. But in FSA post-processing, these trigger nets are eliminated from the set of Trojan nets as their probability feature value ranges in the Free feature range. Additionally, combinational and sequential controllability (with no-scan assumption) values are also less than the average SCOAP value ($\theta_{cc}$ and $\theta_{sc}$) of Free nets.

We can observe from \cref{fig:vipr_framework_merged_seq_func_struct,fig:vipr_framework_merged_seq}, the effect of feature selection when sequential Trojans are used for training the model. The performance of the ML model is reduced when both structural and functional features are used to train the ML model. We believe this reduction in performance is the result of strict structural features, which do not necessarily aid in the identification of these Trojan classes. Additionally, some of the features like distance from flip-flop input, distance from flip-flop outputs, etc., get negatively affected by the inserted sequential Trojans. Using the subset of features identified during feature optimization, the results for sequential training are more acceptable as the feature selection removes the negative bias introduced by some of the structural features. A similar phenomena can be observed when combinational training is used to classify the Trojan nets. We observe the effect of structural features is less impactful for combinational training when compared to structural features in sequential training. However, the results without feature selection for combinational training introduce additional false positives as it becomes difficult to learn the diverse feature space by the ML model. We believe structural features will be more useful for side-channel or always on Trojans.

As shown in \cref{tab:final_voted_results}, the proposed approach can detect suspicious parts in the design with few false positives and false negatives for combinational (C) as well sequential (S) Trojans present in various Trust-HUB designs. The first two columns show the final set of results which are graphically represented by the fourth column of each of the individual designs in Fig.~\ref{fig:vipr_framework_merged_comb_func} and Fig.~\ref{fig:vipr_framework_merged_seq}. The third column is obtained by performing voting across all (3-combinational and 3-sequential types of training database) FSA post-processed VIPR results. In the case of equal votes for Trojan and Free nets, the net is considered a Free net. The fourth column represents results presented in \cite{hoque2018hardware}. Here, we can see that our proposed framework performs better as our approach considers design-specific bias while performing the ML model training. We observed up to 92.85\% reduction in the false-positive numbers with the application of proposed post-processing algorithms. The second to last column represents our implementation of the SC-COTD \cite{sc_cotd} approach. Here, we can observe that the number of false positives is high compared to the proposed approach. By analyzing the features of the nets, we observe that multiple Free nets which are in the vicinity of the Trojan nets have a similar set of feature values. Hence, the usage of only sequential and combinational SCOAP is not enough to detect the Trojan nets. 

\begin{figure}[h]
	\centering
    \includegraphics[width=0.9\linewidth]{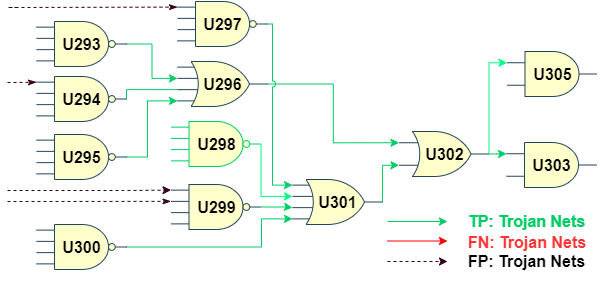}
	\caption{VIPR framework output for RS232-T1000 design with combinational 5-trigger Denial of Service (DoS) Trojan trained ML model } 
	\label{fig:vipr_framework_op}
	
\end{figure}

\cref{fig:vipr_framework_op} shows extracted Trojan circuit for suspicious design RS232-T1000. Here, 5-triggered combinational Trojans are inserted in the suspicious design to create the training data, and the entire framework is run to get the final hyper-graph as shown in Fig.~\ref{fig:vipr_framework_op}. As shown in the figure, all the false positive nets are connected to the original Trojan body. For all the False-negative Trojan nets, their feature values are similar to those of Free nets. Hence, classifying these nets as Trojan nets is not possible. 

Along with the selected features, the effect of remaining set of features was observed on the evaluations of the framework. For example, adding fanin and fanout features in the proposed feature set increased the number of false positives and negatives in ML predictions. Also, we observed that the distribution of structural features for various Trust-Hub Trojan inserted design is not consistent. But these features give a good insight into the design space that can be used by an attacker to insert the Trojan. In future work, these additional features will be used in post-processing routines to remove some of the false positives.

\section{Conclusion}

We have presented VIPR, a hardware Trojan detection method for 3PIP trust verification. We utilize the pseudo-self referencing approach in the hardware Trojan detection, which takes advantage of the design under test to create the training database. This process enables any machine learning model to learn the features of the suspicious design instead of learning features from the database that is generated using different Trojan free designs as discussed in \cite{hoque2018hardware}. We performed a detailed analysis on multiple structural and functional features that can be used to train an ML model. Finally, we introduced three levels of post-processing procedure to further enhance the results by localizing the Trojan body and removing the false positives. We have performed extensive evaluation of VIPR for a set of benchmark designs and Trojan instances, which show promising detection accuracy. 
Future work will involve extension of the feature set to improve the accuracy further, consideration of emerging Trojan attacks, and integrating VIPR into commercial tool flow.

\bibliographystyle{IEEEtran}
\bibliography{IEEEexample}

\end{document}